\documentclass[preprint,floatfix,superscriptaddress,preprintnumbers,showpacs,showkeys]{revtex4}      
\usepackage{amssymb}
\usepackage{amsmath}
\usepackage{amsfonts}
\usepackage{latexsym,graphicx,epsfig}
\usepackage{color}
\usepackage{graphicx}

\newcommand{\eq}{\begin{eqnarray}}
\newcommand{\en}{\end{eqnarray}}

\begin{document}

\preprint{SLAC-PUB-17184}

\title{QCD Constituent Counting Rules for Neutral Vector Mesons}  

\author{Stanley J. Brodsky}
\affiliation{SLAC National Accelerator Laboratory, Stanford University, 
Stanford, CA 94309, USA}

\author{Richard F. Lebed}
\affiliation{Department of Physics, Arizona State University, Tempe, 
Arizona 85287-1504, USA}

\author{Valery E. Lyubovitskij}
\affiliation{Institut f\"ur Theoretische Physik, Universit\"at
T\"ubingen, Kepler Center for Astro and Particle Physics,
Auf der Morgenstelle 14, D-72076, T\"ubingen, Germany}
\affiliation{Departamento de F\'\i sica y Centro Cient\'\i fico
Tecnol\'ogico de Valpara\'\i so-CCTVal,  
Universidad T\'ecnica Federico Santa Mar\'\i a, 
Casilla 110-V, Valpara\'\i so, Chile} 
\affiliation{Department of Physics, Tomsk State University,
634050 Tomsk, Russia}
\affiliation{Laboratory of Particle Physics,
Tomsk Polytechnic University,
634050 Tomsk, Russia}

\begin{abstract}
QCD constituent counting rules define the scaling behavior of
exclusive hadronic scattering and electromagnetic scattering
amplitudes at high momentum transfer in terms of the total number of
fundamental constituents in the initial and final states participating
in the hard subprocess. The scaling laws reflect the twist of the
leading Fock state for each hadron and hence the leading operator that
creates the composite state from the vacuum. Thus, the constituent
counting scaling laws can be used to identify the twist of exotic
hadronic candidates such as tetraquarks and pentaquarks. Effective
field theories must consistently implement the scaling rules in order
to be consistent with the fundamental theory. Here we examine how one
can apply constituent counting rules for the exclusive production of
one or two neutral vector mesons $V^0$ in $e^+ e^-$ annihilation,
processes in which the $V^0$ can couple via intermediate photons.   
In case of a (narrow) real $V^0$, the photon virtuality is fixed to a
precise value $s_1 = m_{V^0}^2$, in effect treating the $V^0$ as a
single fundamental particle.  Each real $V^0$ thus contributes to the
constituent counting rules with $N_{V_0} = 1$. In effect, the leading
operator underlying the $V^0$ has twist 1. Thus, in the specific
physical case of single or double on-shell $V^0$ production via
intermediate photons, the predicted scaling from counting rules
coincides with Vector Meson Dominance (VMD), an effective theory that
treats $V^0$ as an elementary field. However, the VMD prediction
fails in the general case where the $V^0$ is not coupled through an
elementary photon field, and then the leading-twist interpolating
operator has twist $N_{V_0} = 2$. Analogous effects appear in $pp$
scattering processes.
\end{abstract}

\pacs{12.38.Aw,12.40.Vv,13.66.Bc,14.40.Be}
  
\keywords{electron-positron annihilation,
hadron structure, quantum chromodynamics, 
vector meson dominance, electroweak bosons, tetraquarks}

\today

\maketitle

\section{Introduction}
\label{intro}

One of the distinctive consequences of the underlying conformal
features of gauge theories such as QCD is counting rules for hard
exclusive processes.  In such processes, one can factorize the
physical scattering amplitude as the convolution of a hard-scattering
quark and gluon amplitude $T_H$ with the product of hadronic
distribution amplitudes $\phi_H(x,Q)$.  The resulting scaling for the
differential cross section at large momentum transfer
reads~\cite{Brodsky:1973kr,Matveev:1973ra,Lepage:1980fj} $d\sigma/dt
\sim 1/S^{N-2}$, where $S$ is a generic hard scale, and $N = N_i +
N_f$ is the total number of fundamental constituents participating in
the hard subprocess.  The number of constituents of each hadron
entering the scattering amplitude coincides with the number of
particles in its leading Fock state and thus with the twist of the
leading operator that creates the composite state from the vacuum.
For example, the scaling prediction for exclusive cross sections such
as fixed-angle hadron-hadron scattering
is~\cite{Brodsky:1973kr,Matveev:1973ra,Lepage:1980fj}:
\eq\label{eq:count1}
{\frac{d\sigma}{dt}}(A+ B \to C +D) \propto
\frac{F(\theta_{\rm CM})}{S^{N-2}}\,,
\en 
where $N= N_A + N_B + N_C + N_D $ is the total twist or number of
elementary constituents.  When dealing with hadrons, one must take
into account their quark content and use $N_M = 2$ and $N_B = 3$ for
each meson and baryon, respectively.  One also predicts logarithmic
corrections from the behavior of the running couplings entering $T_H$
and the Efremov-Radyushkin-Brodsky-Lepage (ERBL) evolution of the
distribution amplitude.  

The constituent counting rules are completely rigorous {\em when\/}  
they are applied properly.  The leading-twist contribution to the
power-law falloff of a cross section for any exclusive or
semi-inclusive process depends upon the twist of the operators that
couple the hadron to the hard subprocess.  The twist $\tau$ of a
hadron that couples to a hard-scattering subprocess is computed from
the number $N$ of its fundamental constituent fields interacting in
the hard-scattering subprocess (called {\it active\/} in
Ref.~\cite{Sivers:1975dg}), plus $L$, the relative orbital angular
momenta in the contributing hadronic Fock state.  In contrast, the
cross section for hadrons produced through a soft intermediate
state---such as a neutral vector meson $V^0$ produced via its direct
coupling to a photon of finite virtuality, or a hadron produced from
jet fragmentation---does not have increased power-law falloff.

Note also that effective field theories developed to describe hard
hadronic processes must consistently implement the counting rules in
order to be consistent with the underlying fundamental theory of
QCD\@.  AdS/QCD, which allows the calculation of hadronic amplitudes
using light-front holography~\cite{Brodsky:2014yha}, is good example
of such an effective theory.

\begin{figure}[ht]
\begin{center}
  \includegraphics[width=0.3\textwidth]{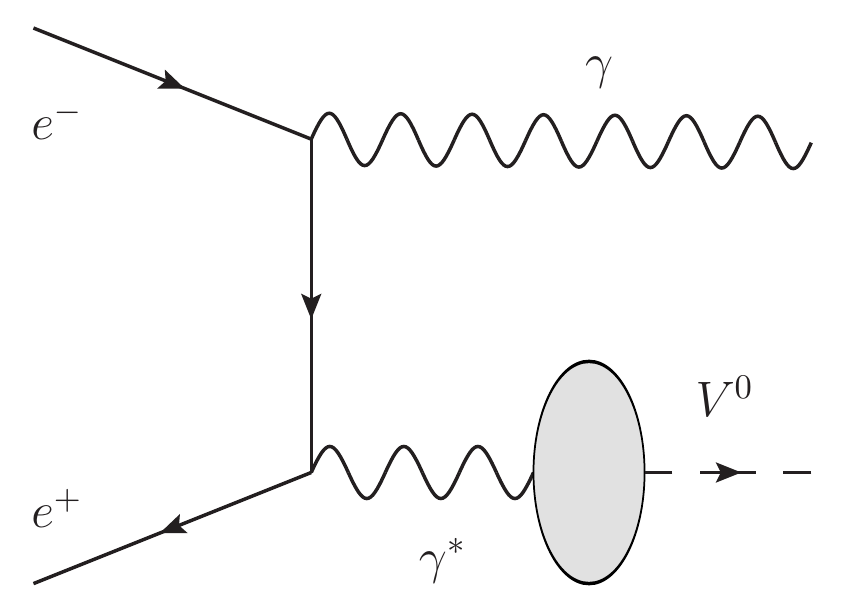}
\end{center}
\caption{Diagram for exclusive production of a vector meson $V^0$ in
  $e^+ e^- \! \to \! \gamma V^0$, the corresponding $u$-channel
  diagram being implied.}
\label{fig:meson}
 \includegraphics[width=0.3\textwidth]{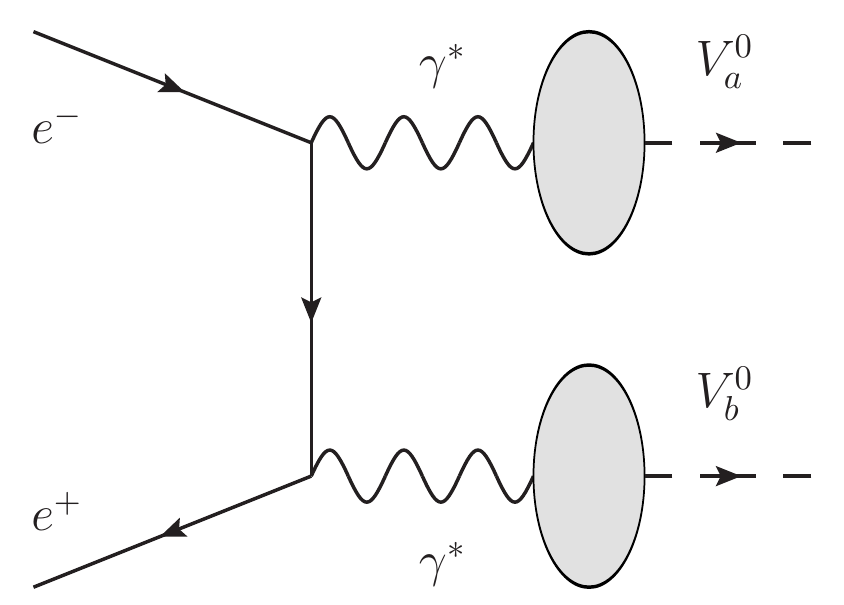}
 \caption{Diagram for exclusive production of vector mesons $V^0_a$
   and $V^0_b$ in $e^+ e^- \! \to \! V^0_a V^0_b$, the corresponding
   $u$-channel diagram being implied.}
\label{fig:2mesons}
\end{figure}

In Refs.~\cite{Brodsky:2015wza,Brodsky:2016uln}, the present authors
studied the application of the constituent counting rules for the
production of tetraquarks, pentaquarks, and $V^0$ in the exclusive
reactions of electroproduction and $p \bar p$ and $e^- e^+$
annihilation.  The purpose of the present paper is to further clarify
our point regarding single and double on-shell $V^0$ production in
$e^+ e^-$ annihilation (see Figs.~\ref{fig:meson} and
\ref{fig:2mesons}), where each $V^0$ couples to the hard
subprocess via a virtual photon. In effect, the leading operator
underlying the $V^0$ has twist $N_{V^0} \! = \! 1$, a point not fully
appreciated in Ref.~\cite{Brodsky:2016uln}. In fact, the possibility
that some of the constituents in a given process counted in the
scaling rule might not be hard is the essence of the critique
of~\cite{Brodsky:2015wza} in Ref.~\cite{Guo:2016fqg}.  Thus, in the
specific physical case of single or double $V^0$ production via
intermediate photons, the predicted scaling from counting rules
coincides with Vector Meson Dominance (VMD)~\cite{Sakurai}, an
effective theory that treats the $V^0$s as elementary fields.  A
modified form of the constituent counting rules therefore holds, and
QCD can be approximated at these exceptional kinematic points by an
effective field theory, the VMD model, which treats $V^0$ as an
elementary field. However, as we shall show, VMD in general is not
consistent with QCD and the constituent counting rules. The VMD
prediction fails in the general case in which the $V^0$ is not coupled
to the hard subprocess via an elementary photon field; in that case,
the leading-twist interpolating operator has twist $N_{V_0} = 2$.

This paper is organized as follows: Section~\ref{sec:Production}
provides general comments about $e^+ e^- \! \to \! \gamma^\ast \gamma$
and $\gamma^\ast \gamma^\ast$ processes, both leptonic and hadronic.
In Sec.~\ref{sec:Example} we examine the process $e^+ e^- \! \to \!
\gamma \, \mu^+ \mu^-$, for which the relevant cross sections have
been explicitly computed, and for which the high-momentum scaling
behavior is explicit, and infer the corresponding behavior for $V^0$
production.  Section~\ref{sec:Incl} shows how the original constituent
counting rules persist in inclusive $e^+ e^-$ processes involving
vector (or scalar or tensor) meson production.  In
Sec.~\ref{sec:ppScat} we consider applications of these ideas in $pp$
scattering processes, and in Sec.~\ref{sec:Concl} we conclude.

\section{Vector-meson production via intermediate photons}
\label{sec:Production}

The most straightforward scaling predicted by the counting rules is
valid in most physical applications, {\it e.g.}, in the
pair-production of mesons, baryons, or tetraquarks in $e^+ e^-$
annihilation~\cite{Brodsky:2015wza}, and occurs whenever all
constituents participate in the hard process, in which cases the scale
$S$ is just Mandelstam $s$, the square of the total center-of-momentum
(c.m.)  energy.  If any of the particles undergo hard scatterings that
constrain them to lie in the forward (beam) c.m.\ direction, the
corresponding factors of $S$ become Mandelstam
$|t|$~\cite{Brodsky:2016uln}.

However, specific physical cases exist, {\it e.g.}, single or double
vector-boson $V^0$ production processes $e^+ e^- \to \gamma V^0$ and
$e^+ e^- \to V^0_a V^0_b$, in which each $V^0$ couples solely to an
intermediate elementary photon field: $\gamma \to V^0$, or a weak
gauge boson: $Z^0 \! \to \! V^0$, $W^\pm \! \to \!  V^\pm$.  In such
cases, the scale associated with the photon virtuality is fixed to a
precise value $s_1 = m_{V^0}^2$, where $m_{V^0}$ is the vector meson
mass.  Therefore, one can treat the $V^0$ (with respect to the
counting rules) as a single fundamental particle, and QCD reduces to
the limit of the Vector Meson Dominance (VMD) model.  In this specific
case, the $V^0$ is approximated by an elementary field with $N_{V^0}
\! = \! 1$ elementary constituents.  Then one has $N-2 =2$ for both
processes $e^+ e^- \to \gamma V^0$ and $e^+ e^- \to V^0_a V^0_b$,
which gives the differential cross-section scaling $d\sigma/dt \propto
1/s^2$, where $s$ is the total c.m.\ energy of lepton pair, or
$1/s|t|$ for forward scattering. This result follows from setting
$s_1 = m_{V^0}^2$ in the $\gamma \to V^0$ transition form factor
$G_V(s_1)$ (calculated, {\it e.g.}, using the soft-wall AdS/QCD
approach) in Ref.~\cite{Brodsky:2016uln}, rather than introducing an
$O(|t|)$ scale in $G_V$ as advocated in that work.  Independently,
this scaling result can be shown explicitly by considering the related
process $e^+ e^- \! \to \! \gamma \, \mu^+ \mu^-$ at high energy but
small invariant mass for the $\mu^+ \mu^-$ pair
(Sec.~\ref{sec:Example}), an exercise that is instructive in
explicitly indicating where the various momentum scales appear.  Let
us stress again that the scaling of the differential cross section
$d\sigma/dt \propto 1/s^2$ in the particular processes of single or
double vector-boson $V^0$ production does not violate the constituent
counting rules, because the exclusive $\gamma \to V^0$ transition
necessarily implies soft QCD, leading one to approximate the $V^0$
(with respect to hard scales) as an elementary field.  In other words,
the presence of soft QCD vertices in hard processes leads to a
decrease of the scaling power in the corresponding differential cross
section by identifying each softly produced hadron composed of $N_a$
constituents with an elementary field: $N_a \to 1$.

Note that the production of a $V^0$ via a photon can be a subleading
contribution to the matrix element of a hard process.  An example of
such process is $V^0$ production in the reaction $e^+ e^- \to V^0
P^0$, where $P^0$ is a neutral pseudoscalar meson ({\it e.g.}, $\pi^0
\! , \eta, \eta^\prime$).  In Ref.~\cite{Guo:2016fqg}, VMD was the
mechanism proposed for the $\gamma \to V^0$ transition in such
processes.  It is clear that this subprocess is $O(\alpha_{\rm
em})$-suppressed in comparison with the leading QCD diagram discussed
in Ref.~\cite{Brodsky:1981kj} for direct production of a $V^0 P^0$
pair by a hard photon, $\gamma^\ast \to V^0 P^0$. As was shown in
Ref.~\cite{Brodsky:1981kj}, the matrix element for $e^+ e^- \to V^0
P^0$ contains a helicity-flip transition form factor $F_{\gamma^\ast
V^0P^0}(s)$, which encodes violation of hadron helicity selection
rules and scales as $1/s^2$ at large $s$. As result, the corresponding
cross section scales as $d\sigma/dt \propto 1/s^5$, {\it i.e.}, it has
an additional $1/s$ falloff compared to helicity-favored modes of
two-meson production ($\pi^+\pi^-$, $K^+K^-$, {\it etc.})  The
mechanism for the $e^+ e^- \to V^0 P^0$ reaction considered in
Ref.~\cite{Guo:2016fqg} gives $d\sigma/dt \propto 1/s^3$, but, as
stressed above, is suppressed by a power of $\alpha_{\rm em}$ in
comparison with the leading QCD diagram.

\section{Lessons from the process
$e^+ e^- \! \to \!  \gamma \, \mu^+ \mu^-$} \label{sec:Example}

\begin{figure}[ht]
\begin{center}
  \includegraphics[width=0.3\textwidth]{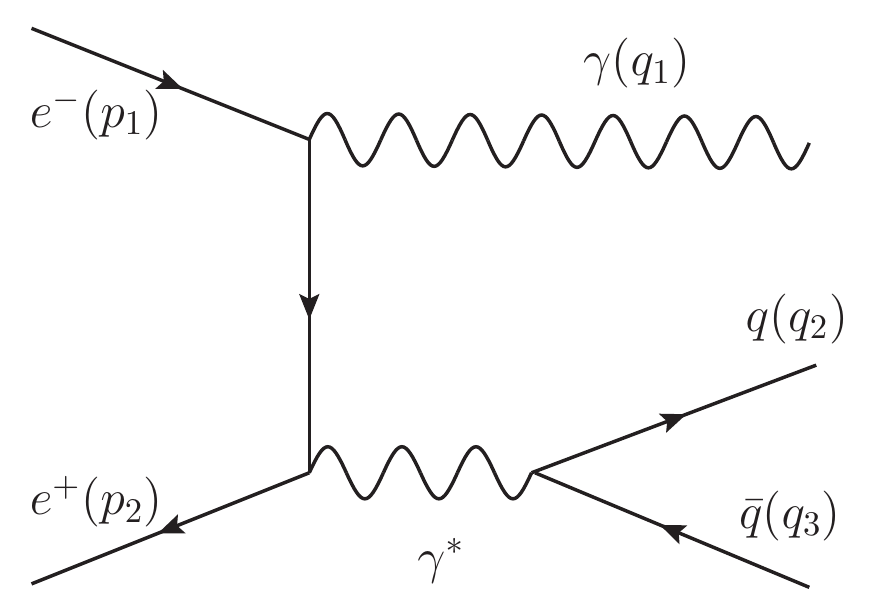}
\end{center}
\caption{Diagram contributing to $e^+ e^- \! \to \! \gamma \gamma^* \!
  \to \! \gamma q \bar q$, the corresponding $u$-channel diagram being
  implied.}
\label{fig:parton}       
\end{figure}

In order to verify or falsify the claim from
Ref.~\cite{Brodsky:2016uln} that the $\gamma^* V^0$ transition form
factor $G_V(q^2)$ scales as $1/\sqrt{|t|}$ for forward scattering in
$e^+ e^- \! \to \! \gamma V^0$, one may study the related process $e^+
e^- \! \to \! \gamma \, \mu^+ \mu^-$, which has been considered for
decades~\cite{Berends:1980yz,Kuraev:1977rp} as a background to $e^+
e^- \! \to \! \mu^+ \mu^-$, and more recently~\cite{Arbuzov:2004gy} in
the {\it initial-state radiation} process, in which the real photon is
hard but the $\mu^+ \mu^-$ pair is soft.  Indeed, the result of
Ref.~\cite{Arbuzov:2004gy} was used to estimate the yield of true
muonium ($\mu^+ \mu^-$) atoms in the process $e^+ e^- \! \to \! \gamma
+ (\mu^+ \mu^-)$~\cite{Brodsky:2009gx}.  The process assumes the same
topology as Fig.~\ref{fig:parton}, with $q\bar q$ replaced by
$\mu^-\mu^+$.

To serve as an orientation, we exhibit the textbook
result~\cite{Peskin:1995ev} of the Born-level cross section for the
pair-annihilation process $e^+ e^- \! \to \! \gamma \gamma$.  One
finds, neglecting masses, and in the forward direction ($m_e^2 \! \ll
\! |t| \! \ll \!  s$),
\begin{equation}
\frac{d\sigma}{dt} \to \frac{2\pi \alpha^2}{s^2 |t|} \frac{s^2 + 2st +
2t^2}{s+t} \to \frac{2\pi \alpha^2}{s|t|} \,
\end{equation}
in agreement with the prediction of Eq.~(\ref{eq:count1}) with $N \!
= \! 4$ and one forward (fermion) propagator.

The full Born-level cross section for $e^+ e^- \!
\to \! \gamma + (\mu^+ \mu^-)$, as can be seen in Eq.~(14) of
Ref.~\cite{Kuraev:1977rp}, possesses a second-order pole $1/(1 \pm
z)^2$ in $z \! \equiv \! \cos \theta_+$, where $\theta_+$ is the
$\mu^+ e^-$ angle in the $e^+ e^-$ c.m.\ frame.  The corresponding
differential cross section $d\sigma/dt$ therefore has separate terms
scaling as $1/|t|^2$ and as $1/|u|^2$.  They clearly arise through the
near-collinear kinematics in which $\mu^+$, $\mu^-$, and $\gamma$ all
lie close to the beam axis but have large relative momenta; in that
case, both the fermion and photon propagators in Fig.~\ref{fig:parton}
contribute the large momentum-transfer factors.

But now restrict to the kinematics of Ref.~\cite{Arbuzov:2004gy}, in
which the momentum transfer $s_1$ of the $\mu^+ \mu^-$ pair is small;
in the exclusive $q\bar q$ case, $s_1 \! = \!  m_{V^0}^2$ (labeled
$q^2$ in Ref.~\cite{Brodsky:2016uln}).  The $e^+ e^- \! \to \! \gamma
+ (\mu^+ \mu^-)$ forward differential cross section then reads
\begin{equation} \label{eq:Smalls1}
\frac{d\sigma}{dt} = \frac{\alpha^3}{s|t|s_1} \left( 2\delta + 1 -
2x_- x_+ \right) dx_- ds_1 \, ,
\end{equation}
where $\delta \! \equiv \! m_{\mu}^2 /s_1$, $x_{\pm} \! \equiv \!
E_{\mu^{\pm}} / (\sqrt{s}/2)$ are the fractional $\mu^\pm$ energies,
and $m_e \! \to \! 0$.  The question then becomes how much the
remaining integrals, those over $dx_-$ and $ds_1$, influence the full
high-momentum scaling of $d\sigma/dt$.  One easily finds that
\begin{equation}
s_1 \to 2s \sin^2 \frac{\epsilon}{2} \, , \ \ \ t \to -2s \sin^2
\frac{\theta}{2} \, ,
\end{equation}
where $\epsilon$ is the $\mu^+ \mu^-$ angle and $\theta$ is the $e^-
\gamma$ angle in the c.m.\ frame.  We are therefore interested in the
hierarchy $s_1 \! \ll \! |t| \! \ll \! s$, or $\epsilon \! \ll \!
\theta \! \ll \! 1$.  Immediately one sees that small $s_1$ requires a
small c.m.\ angle $\epsilon$ between the $\mu^+ \mu^-$ pair; however,
at this stage no similar restriction requires the $\mu^+$ and $\mu^-$
to share their total energy $\sqrt{s}/2$ equally, so that $x_- \! = 1
- x_+$ can assume any value $\in [0,1]$.

We now turn to the exclusive hadronic case ($\mu^- \mu^+ \! \to \!
q\bar q$), in which Eq.~(\ref{eq:Smalls1}) is modified via
multiplication by a color factor 3 and the $\gamma^\ast V^0$
transition form factor $|G_V|^2$.  Here, one may naively think that
the constraint of forming a bound state---that the momenta of the
initial $q\bar q$ pair differ by no more than $O(\sqrt{s_1}) \! = \!
O(\Lambda_{\rm QCD})$---forces their energies to be almost equal when
compared to their total energy $\sqrt{s}/2$, thus forcing $x_- \!
\simeq x_+$ and suppressing the region of support of the $x_-$
integral.  However, this momentum constraint applies to the quarks in
their own c.m.\ frame, but their relative momentum when evaluated in
the $e^+ e^-$ c.m.\ frame must be multiplied by a relativistic boost
factor $\gamma(v) v \! \approx 1/2
\cdot \sqrt{s/s_1}$.  The whole [0,1] interval for the $x_-$ integral
therefore contributes to hadronic bound states.

One is therefore left to consider the $s_1$ integral.  Strictly
speaking, the allowed range of $s_1$ for a vector state $V^0$ of
narrow width is vanishingly small, and $|G_V|^2$ assumes the form of a
decay constant $F_{V^0}^2$ (of dimension mass squared) times a delta
function $\delta(s_1 - m_{V^0}^2)$: Requiring the virtual photon in
Fig.~\ref{fig:parton} to produce only a single exclusive state $V^0$
of squared mass $s_1$ fixes the photon virtuality precisely to equal
$s_1$.  However, the same result obtains if $G_V(s_1)$ is replaced by
a properly normalized Breit-Wigner distribution representing a wide
state such as $\rho^0$.  The form factor $|G_V|^2$ must also decrease
with $s_1$ in order to satisfy unitarity, but this decrease merely
indicates that couplings to highly excited $V^0$'s must decrease with
$s_1 \! = \! m_{V^0}^2$ in order to sum to a finite total.  In the
AdS/QCD calculation of Ref.~\cite{Brodsky:2016uln}, this dependence in
terms of the AdS/QCD scale parameter $\kappa$ would read
$\kappa^2/s_1$.  The expected ``large'' momentum-scale suppression in
$G_V$ in exclusive transitions due to constituent counting rules
actually comes from $s_1$.

Knowledge of the larger scale $|t|$ by the $V^0$ is lost in the
propagation of the photon.  For the most naive form of the counting
rules to hold, all propagators and fermion couplings must contribute
large scales to the amplitude, and the virtual photon in this case
contributes only $1/s_1$.  One concludes that the forward cross
section for a {\em strictly} two-body process $e^+ e^- \! \to \!
\gamma V^0$ in which $V^0$ contains two fundamental constituents [$N
\! = \! 5$ in Eq.~(\ref{eq:count1})] should scale as $d\sigma/dt \!
\sim \!  (\alpha^3 / s|t|) |G_V(s_1)|^2$, with $G_V(s_1) \! \sim \!
1/\sqrt{s_1}$.

In contrast, Ref.~\cite{Brodsky:2016uln} concluded that $G_V(s_1) \!
\sim \!  1/\sqrt{|t|}$, the large scale $|t|$ reemerging through a
hard-gluon exchange needed to bind the otherwise noncollinear pair
$q\bar q$ into the bound state $V^0$.  However, as noted above, the
formation of a single photon of virtuality $s_1 \! \ll \!  |t|$
completely erases the system's memory of the large scale $|t|$:
Emission of a hard gluon in this case is not natural.  Moreover, even
though the $q\bar q$ pair can have $O(\sqrt{s})$ energies in the c.m.,
their momentum invariants (their masses and $s_1$) are small.

Consider instead a process such as that illustrated in
Fig.~\ref{fig:parton}, except that the photon virtuality $s_1$ does
not precisely equal $m_{V^0}^2$, but rather assumes a value of
$O(|t|)$ (because the process is still one of forward scattering).
The inclusive process $e^+ e^- \! \to \! \gamma q\bar q$ has a much
greater phase space than does the exclusive process $e^+ e^- \! \to \!
\gamma V^0$, and its cross section scales in the forward direction as
$d\sigma/dt \! \sim \!  1/s|t|^2$ (as seen above for $e^- e^+ \! \to \!
\gamma \mu^+ \mu^-$).  This inclusive rate does indeed include a
portion of the exclusive channel $e^+ e^- \! \to \! \gamma V^0$, but
only from the large-$|t|$ tail of the $V^0$ line-shape.  It also
includes contributions from $e^+ e^- \! \to \! \gamma V^0$ {\em plus
additional soft hadrons\/} such that the total hadronic system has
invariant mass-square of $O(|t|)$, which can be misidentified as the
exclusive channel $e^+ e^- \! \to \!
\gamma V^0$ if the soft hadrons escape detection.

In summary, the correct high-momentum forward-angle scaling for the
genuine two-body exclusive $e^+ e^- \! \to \! \gamma V^0$ cross
section  is $d\sigma/dt \! \sim \! 1/s|t|$, rather than
$1/s|t|^2$ as given in Ref.~\cite{Brodsky:2016uln}.  However, tails of
the original process and processes that can be misidentified as $e^+
e^- \! \to \! \gamma V^0$ give contributions scaling as $1/s|t|^2$;
and since they have much greater available phase space, they may
dominate the observed rate of $e^+ e^- \! \to \! \gamma V^0$ even if
$|t|$ is rather larger than $m_{V^0}^2$.  Completely analogous
comments hold for the process $e^+ e^- \! \to \!  V^0_a V^0_b$.

\section{Inclusive $e^+ e^-$ processes with
virtual photons vs.\ VMD} \label{sec:Incl}

When the $V^0$s are off-shell, an extra power falloff in the large
scale appears for each meson state. The forward-scattered virtual
photons then carry $O(|t|)$ momentum transfers, and the constituent
counting rules read~\cite{Brodsky:2016uln}:
\eq
\frac{d\sigma}{dt}(e^+ e^- \to \gamma V^{*0}) \sim \frac{1}{s^2 |t|^2} \,
, \quad \frac{d\sigma}{dt}(e^+ e^- \to V^{*0}_a V^{*0}_b) \sim
\frac{1}{s^2 |t|^3}\,.
\en
Crucially, the full QCD theory differs from an effective field theory
developed from VMD in their applications to physical processes.  In
particular, VMD makes no distinction between on-shell and off-shell
$V^0$, which leads to an incorrect off-shell behavior of the
$\gamma^\ast \to V^0$ transition.  In Ref.~\cite{Brodsky:2016uln} we
explicitly showed that a constant value for this transition is ruled
out by the nontrivial form factor $G_V(q^2)$, where $q$ is the photon
(or $V^0$) four-momentum.  In fact, $G_V \sim 1/|t|^{1/2}$ at large $t
= q^2$, consistent with perturbative QCD (pQCD) and constituent
counting rules.

\begin{figure}[ht]
\begin{center}
  \includegraphics[width=0.5\textwidth]{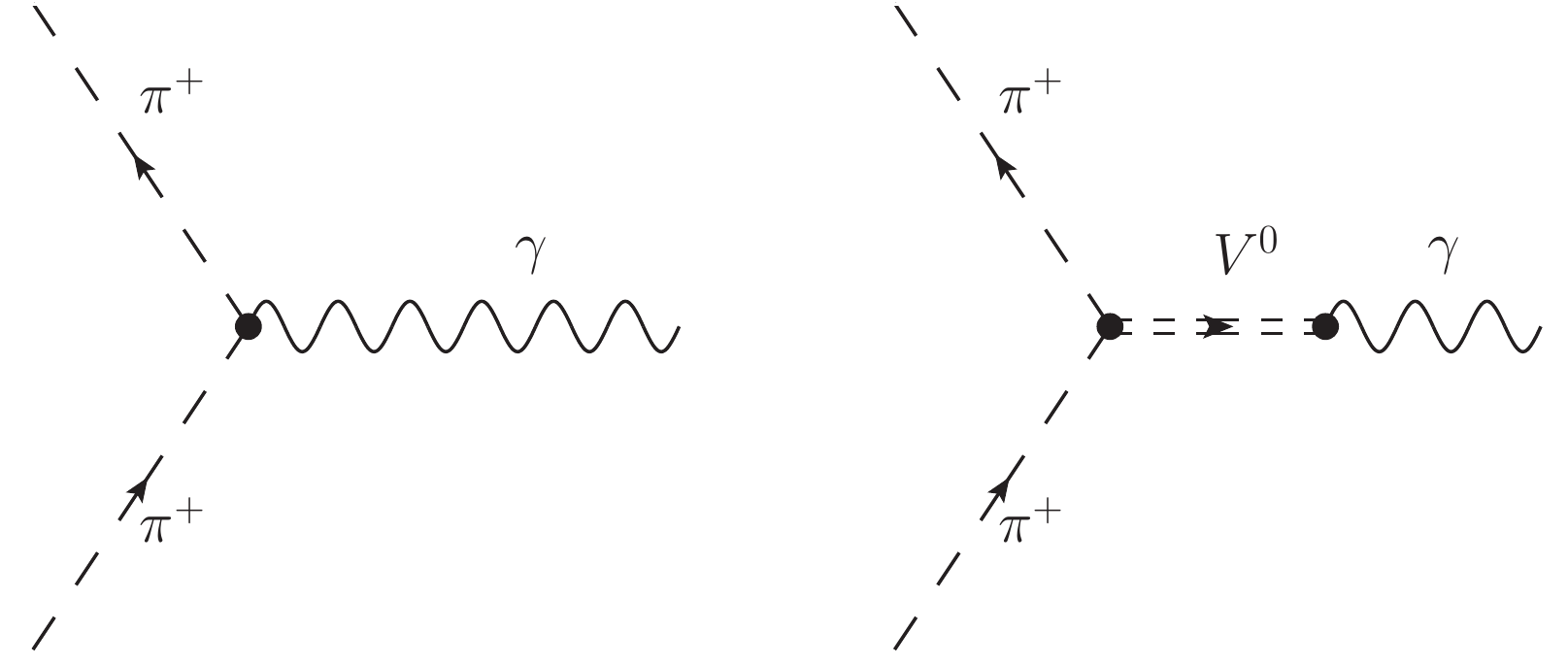}
\end{center}
\caption{Diagrams (contact and vector-meson exchange) contributing to
the electromagnetic form factor of the pion in effective field
theories involving vector mesons as elementary fields.}
\label{fig:pion}       
\end{figure}

In Ref.~\cite{Brodsky:2016uln}, we pointed out another property
distinguishing pQCD from effective field theories treating $V^0$ as
elementary fields, {\it i.e.}, in its application to electromagnetic
form factors of hadrons.  Let us review this important point: In
effective theories (like VMD~\cite{Sakurai}, chiral perturbation
theory~\cite{Ecker:1989yg}, or the hidden-symmetry approach with
vector mesons as dynamical gauge bosons~\cite{Bando:1987br}) that
treat the $V^0$ as an elementary field, the form factor $G_V(q^2)$ is
a constant.  One obtains different scaling contributions of the
relevant diagrams with elementary $V^0$ and with pQCD. For example,
in the case of the pion electromagnetic form factor $F_\pi(Q^2)$, one
may split the VMD result into the contact and the vector-meson
exchange diagrams (Fig.~\ref{fig:pion}).  The contact diagram gives 1
(a constant contribution at large $Q^2 = - q^2$), whereas the
vector-meson exchange diagram gives a $(-1 + m_{V^0}^2/Q^2)$
contribution.  Summing, one arrives at a $m_{V^0}^2/Q^2$ scaling.  In
contrast, using pQCD counting rules, the contact diagram turns out to
be of leading order ($1/Q^2$), whereas the $V^0$-exchange diagram is
subleading [$(1/Q^2)^{3/2}$] at large $Q^2$.  Of course, at fixed $Q^2
= m_{V^0}^2$ the contributions do not cleanly separate.  Another
crucial point is that, while the scaling of $F_\pi(Q^2)$ in VMD is
formally $m_{V^0}^2/Q^2$ due to the $V^0$ propagator, this factor has
no connection with $1/Q^2$ scaling in pQCD due to hard-gluon exchanges
between constituent quarks in the pion.  Therefore, the $1/Q^2$
falloff of $F_\pi(Q^2)$ in VMD coincides with pQCD accidentally, and
is not related to the physical nature of strong interactions at high
scales.  As a consequence, an effective field theory of VMD completely
fails in the description of the electromagnetic form factors of
baryons and multi-constituent hadronic systems (tetraquarks,
pentaquarks, {\it etc.}) at high scales.  Without pQCD it is
impossible to produce the $1/Q^{2 (N-1)}$ falloff of the
electromagnetic form factors of hadrons containing $N$ constituents.
Note that a criticism of VMD in the description of data for
photon-hadron interactions at high energies was also stressed in
Ref.~\cite{Friedman:1991nq}.  In particular, 
\cite{Friedman:1991nq} argues that VMD is not a suitable framework for
a description of deep-inelastic scattering over the full kinematical
range.

Additionally, in the generic case of exclusive on-shell hadron
production, it is not possible to approximate the hadrons as
elementary fields.  In the recent paper Ref.~\cite{Chumakov:2017kkf},
the exclusive production processes of scalar $S = 0^{++}$ and tensor
$T = 2^{++}$ mesons through single-photon annihilation $e^+ e^- \to
\gamma^* \to \gamma \, S(T)$ were analyzed.  Here, the transition form
factors of $\gamma^* \to \gamma S$ and $\gamma^* \to \gamma T$ are not
constants, and scale as $F_{\gamma^\ast \! \gamma S}(s) \sim 1/s$ and
$F_{\gamma^\ast \!  \gamma T}(s) \sim 1/s^2$ at large $s$, consistent
with the scaling of the corresponding form factors at large values of
virtual-photon squared Euclidean momentum $Q^2$: $F_{\gamma^\ast \!
\gamma S}(Q^2) \sim 1/Q^2$~\cite{Kroll:2016mbt} and $F_{\gamma^\ast \!
\gamma T}(Q^2) \sim 1/Q^4$~\cite{Achasov:2015pha}.  The scaling of the
form factors follows directly from using the differing twist counting
for the $S$- and $T$-creating operators~\cite{Brodsky:1981kj}.  As a
result, both differential cross sections scale as
$\frac{d\sigma}{dt}(e^+ e^- \to \gamma + S(T)) \sim 1/s^3$, in
agreement with constituent-quark counting rules that treat real scalar
and tensor mesons as $q\bar q$ systems with $N_{S(T)} = 2$ substituted
into the counting formula~(\ref{eq:count1}) for $d\sigma/dt$.
Consistent with Ref.~\cite{Brodsky:2015wza}, when scalar and tensor
mesons are considered as tetraquark systems of two tightly bound color
diquarks, the corresponding transition form factors and differential
cross sections have the same falloffs as in the $q\bar q$ case.  For
other tetraquark or two-hadron molecular configurations, the
transition form factors $F_{\gamma^\ast \! \gamma S(T)}(s)$ and the
differential cross section $d\sigma/dt$ have additional falloffs
scaling as $1/s$ and $1/s^2$, respectively.

Again, we point out that the case of single and double neutral
vector-meson production via intermediate photon or the weak gauge
boson fields is very specific, constraining $V^0$ (with respect to the
counting rules) to acting as effectively fundamental (structureless)
particles; it is the result of an exceptional case in which some of
the internal propagators ({\it i.e.}, virtual gauge bosons) are
explicitly excluded from carrying large off-shell virtuality.

\section{Vector-meson production in $p p$ scattering}
\label{sec:ppScat}

In this section we discuss $V^0$ production in $p p$ scattering
processes.  The hadronic angular momentum dependence of hard exclusive
QCD processes is controlled by the Brodsky-Lepage helicity selection
rules~\cite{Brodsky:1981kj,Brodsky:1989pv}, which state that the total
hadron helicity is conserved from the initial to the final state, up
to higher-twist corrections appearing as inverse powers of the hard
scale.  This result was used for $e^+ e^- \to V^0 P^0$ in
Sec.~\ref{sec:Production}.  Taking here $V^0 \! = \rho^0$, we consider
three specific cases of {\em semi-inclusive\/} $\rho^0$-meson
production~\cite{Sivers:1975dg,Brodsky:1998sr,Arleo:2009ch}: (1) the
reaction $p p \to \rho^0 X$, with the $\rho^0$ produced from jet
fragmentation and $X$ being any hadrons; (2) the reaction $p p \to
\rho^0_D X$, with a ``direct'' $\rho^0_D$ produced at high $p_T$ in
isolation from other hadrons on the trigger side ({\it i.e.}, without
any same-side particles); and (3) the reaction $p p \to \gamma^\ast X
\to \rho^0_D X$, where a single virtual photon produces a ``direct''
$\rho^0$, which again is isolated on the trigger side.

Reaction (1) has normal conformal scaling (modulo log corrections).
Consistent with Eq.~(\ref{eq:count1}), the differential cross section
for semi-inclusive production of a single hadron $\rho^0$ with form
factor $F$ scales as
\eq 
\frac{d\sigma}{d^3p/E} \sim \frac{F(x_T)}{p_T^4}\,, \label{eq:Leading}
\en
where $p_T$ and $x_T = {2p_T/\sqrt s}$ are the transverse momentum and
its light-cone fraction, respectively.

In case (2) the $\rho^0_D$ couples via a $q \bar q$ to the hard
underlying hadron subprocess.  The corresponding differential cross
section then has an additional $1/p_T^2$ falloff in comparison with
case (1) and scales as
\eq
\frac{d\sigma}{d^3p/E} 
\sim \frac{F(x_T)}{p_T^6}\,,
\en
reflecting the corresponding twist-2 operator and the $|q \bar
q\rangle $ Fock state of the $\rho^0$.  Note that reaction (2) is
power-suppressed at high $p_T$ (being higher twist), but the $\rho^0$
in this case exhibits color transparency~\cite{Brodsky:2008qp}: It is
produced directly from the hard subprocess as a small-sized
color-singlet state and can propagate through a nuclear medium with
minimal interactions.  In contrast to reaction (2), the process $p p
\to \gamma X$ with an isolated photon occurs at leading twist since
the photon can couple directly to the hard process without additional
power suppression.

Consideration of $p p \to \gamma X$ leads to case (3), which again
scales like $\frac{d\sigma}{d^3p/E} \sim 1/p_T^4$ as in
Eq.~(\ref{eq:Leading}), since in this case the $\rho^0$ couples softly
via the twist-1 photon field to the hard subprocess without an
additional power suppression.  Reaction (3) exhibits the same type of
behavior as discussed Secs.~\ref{sec:Production} and \ref{sec:Example}
for $e^+ e^- \to \gamma V^0$ and $e^+ e^- \to V^0_a V^0_b$.

\section{Conclusions} \label{sec:Concl}

Let us summarize the main results of this paper.  We examined the
application of QCD constituent counting rules to exclusive processes
involving neutral vector mesons $V^0$.  In particular, we considered
exclusive production of one or two $V^0$ via intermediate photons from
$e^+ e^-$ annihilation, and in $pp$ scattering.  In case of a real
$V^0$, the photon virtuality $s_1$ can be fixed to a precise value
$m_{V^0}^2$, in effect treating the $V^0$ as a single fundamental
particle.  Therefore, each real $V^0$ contributes to the constituent
counting rules with $N_{V_0} = 1$.  Because the leading operator
underlying the $V^0$ has twist~1, in the case of single or double
on-shell $V^0$ production via intermediate photons, the predicted
scaling from counting rules coincides with Vector Meson Dominance
(VMD), an effective theory that treats vector mesons as elementary
fields.

However, the VMD prediction fails in the general case where the $V^0$
is not coupled solely through an elementary photon field, and in that
case the leading-twist interpolating operator has $N_{V_0} = 2$.
Furthermore, VMD fails in the case of off-shell coupling of the
electromagnetic field with hadrons at large momentum scales because
this approach, by construction, does not respect the constituent
structure of hadrons and hard-gluon exchange at large scales.  As a
result, the large-$Q^2$ scaling of electromagnetic form factors of
hadrons with $N \ge 3$ constituents in VMD is not consistent with that
from pQCD. Only in the case of $q\bar q$ systems (conventional
mesons) is the VMD prediction of $1/Q^2$ scaling formally similar to
that of pQCD because of the $1/Q^2$ behavior of the $V^0$ propagator.
One should also note the criticism of VMD in the description of data
for photon-hadron interactions at high energies stressed before in
Ref.~\cite{Friedman:1991nq}.

\begin{acknowledgments}
  We thank Feng-Kun Guo for suggesting the examination of the process
  $e^+ e^- \! \to \! \gamma \mu^+ \mu^-$.  This research was supported
  by the U.S.\ Department of Energy, contract DE--AC02--76SF00515
  (SJB), by the U.S.\ National Science Foundation under Grant No.\
  PHY-1403891 (RFL), by the German Bundesministerium f\"ur Bildung und
  Forschung (BMBF) under Project 05P2015 - ALICE at High Rate
  (BMBF-FSP 202): ``Jet and fragmentation processes at ALICE and the
  parton structure of nuclei and structure of heavy hadrons'', by
  CONICYT (Chile) PIA/Basal FB0821, by Tomsk State University
  Competitiveness Improvement Program, by the Russian Federation
  program ``Nauka'' (Contract No.\ 0.1764.GZB.2017), and by Tomsk
  Polytechnic University Competitiveness Enhancement Program (Grant
  No. VIU-FTI-72/2017) (VEL).  SLAC-PUB-17184.
\end{acknowledgments}


\begin{thebibliography}{}
%
\bibitem{Brodsky:1973kr}
  S.J.~Brodsky and G.R.~Farrar,
  Phys.\ Rev.\ Lett.\  {\bf 31}, 1153 (1973).
%
\bibitem{Matveev:1973ra}
  V.A.~Matveev, R.M.~Muradian, and A.N.~Tavkhelidze,
  Lett.\ Nuovo Cim.\  {\bf 7}, 719 (1973).
%
\bibitem{Lepage:1980fj}
  G.P.~Lepage and S.J.~Brodsky,
  Phys.\ Rev.\ D {\bf 22}, 2157 (1980).
%
\bibitem{Sivers:1975dg} 
  D.W.~Sivers, S.J.~Brodsky, and R.~Blankenbecler,
  Phys.\ Rept.\  {\bf 23}, 1 (1976).
%
\bibitem{Brodsky:2014yha}
  S.J.~Brodsky, G.F.~de Teramond, H.G.~Dosch, and J.~Erlich,
  Phys.\ Rept.\  {\bf 584}, 1 (2015)
  [arXiv:1407.8131 [hep-ph]].
%
\bibitem{Brodsky:2015wza}
  S.J.~Brodsky and R.F.~Lebed, 
  Phys.\ Rev.\ D {\bf 91}, 114025 (2015)
  [arXiv:1505.00803 [hep-ph]].
%
\bibitem{Brodsky:2016uln} 
  S.J.~Brodsky, R.F.~Lebed, and V.E.~Lyubovitskij,
  Phys.\ Lett.\ B {\bf 764}, 174 (2017)
  [ar\-Xiv:1609.06635 [hep-ph]].
%
\bibitem{Guo:2016fqg} 
  F.-K.~Guo, U.-G.~Mei{\ss}ner, and W.~Wang,
  Chin.\ Phys.\ C {\bf 41}, 053108 (2017)
  [arXiv:1607.04020 [hep-ph]].
%
\bibitem{Sakurai}J.J.~Sakurai, {\it Currents and Mesons},  
 University of Chicago Press, Chicago, 1969. 
%
\bibitem{Brodsky:1981kj} 
  S.J.~Brodsky and G.P.~Lepage,
  Phys.\ Rev.\ D {\bf 24}, 2848 (1981).
%
\bibitem{Berends:1980yz} 
  F.A.~Berends and R.~Kleiss,
  Nucl.\ Phys.\ B {\bf 177}, 237 (1981).
%
\bibitem{Kuraev:1977rp} 
  E.A.~Kuraev and G.V.~Meledin,
  Nucl.\ Phys.\ B {\bf 122}, 485 (1977).
%
\bibitem{Arbuzov:2004gy} 
  A.B.~Arbuzov, E.~Bartos, V.V.~Bytev, E.A.~Kuraev, and Z.K.~Silagadze,
  JETP Lett.\  {\bf 80}, 678 (2004)
  [Pisma Zh.\ Eksp.\ Teor.\ Fiz.\  {\bf 80}, 806 (2004)].
%
\bibitem{Brodsky:2009gx} 
  S.J.~Brodsky and R.F.~Lebed,
  Phys.\ Rev.\ Lett.\  {\bf 102}, 213401 (2009) 
  [arXiv:0904.2225 [hep-ph]].
%
\bibitem{Peskin:1995ev} 
  M.E.~Peskin and D.V.~Schroeder,
  {\em An Introduction to Quantum Field Theory},
  Westview Press, New York (1995).
%
\bibitem{Ecker:1989yg} 
  G.~Ecker, J.~Gasser, H.~Leutwyler, A.~Pich, and E.~de Rafael,
  Phys.\ Lett.\ B {\bf 223}, 425 (1989).
%
\bibitem{Bando:1987br} 
  M.~Bando, T.~Kugo, and K.~Yamawaki,
  Phys.\ Rept.\  {\bf 164}, 217 (1988).
%
\bibitem{Friedman:1991nq} 
  J.~I.~Friedman,
  Rev.\ Mod.\ Phys.\  {\bf 63}, 615 (1991).
%
\bibitem{Chumakov:2017kkf} 
  A.G.~Chumakov, T.~Gutsche, V.E.~Lyubovitskij, and I.~Schmidt,
  Phys.\ Rev.\ D {\bf 96}, 094018 (2017). 
  [arXiv:1709.07234 [hep-ph]].
%
\bibitem{Kroll:2016mbt}
  P.~Kroll,
  Eur.\ Phys.\ J.\ C {\bf 77}, 95 (2017) 
  [arXiv:1610.01020 [hep-ph]]
%
\bibitem{Achasov:2015pha}
  N.N.~Achasov, A.V.~Kiselev, and G.N.~Shestakov,
  JETP Lett.\  {\bf 102}, 571 (2015)
  [Pisma Zh.\ Eksp.\ Teor.\ Fiz.\  {\bf 102}, 655 (2015)]  
  [arXiv:1509.09150 [hep-ph]].
%
\bibitem{Brodsky:1989pv} 
  S.J.~Brodsky and G.P.~Lepage,
  Adv.\ Ser.\ Direct.\ High Energy Phys.\  {\bf 5}, 93 (1989).
%
\bibitem{Brodsky:1998sr} 
  S.J.~Brodsky, M.~Diehl, P.~Hoyer, and S.~Peigne,
  Phys.\ Lett.\ B {\bf 449}, 306 (1999)
  [hep-ph/9812277];
%
\bibitem{Arleo:2009ch}
  F.~Arleo, S.J.~Brodsky, D.S.~Hwang, and A.M.~Sickles,
  Phys.\ Rev.\ Lett.\  {\bf 105}, 062002 (2010)
  [arXiv:0911.4604 [hep-ph]]. 
%
\bibitem{Brodsky:2008qp} 
  S.~J.~Brodsky and A.~Sickles,
  Phys.\ Lett.\ B {\bf 668}, 111 (2008)
  [arXiv:0804.4608 [hep-ph]]. 

\end{thebibliography}
\end{document}